\documentclass[epsf,12pt]{article}
\usepackage{amssymb}
\begin{document}
\baselineskip2em
\begin{center}
{\large \bf Interface states in  junctions of two semiconductors
with intersecting dispersion curves}

\bigskip

{A.~V.~Kolesnikov$^*$, R.~Lipperheide$^\dagger$,
A.~P.~Silin$^\ddagger$ and  U.~Wille$^\dagger$}
\bigskip

{\it $^*$ Ruhr-University Bochum, Universit\"atstr. 150, D-447801, Bochum, Germany} 

{\it $^\dagger$ Bereich Theoretische Physik, Hahn-Meitner-Institut
Berlin, \\ D-14091 Berlin, Germany}

{\it $^\ddagger$ Tamm Theoretical Department of the Lebedev Physical
Institute,~RAS, \\  Leninskii pr. 53, 117924, Moscow,  Russia}
\bigskip
\end{center}

\vspace{1.2cm}



junctions, and heterojunctions.

\vspace{0.8cm}

{\small {\bf Abstract.}
-- A novel type of shallow interface state in junctions of two semiconductors
without band inversion is identified within the envelope function
approximation, using the two-band model.  It occurs in abrupt junctions when
the interband velocity matrix elements of the two semiconductors differ and the
bulk dispersion curves intersect.  The \linebreak in-plane dispersion of the
interface state is found to be confined to a finite range of momenta centered
around the point of intersection.  These states turn out to exist also in
graded junctions, with essentially the same properties as in the abrupt case.}

\newpage
The rapid progress in the synthesis of mesoscopic structures during the last
decades has evoked renewed attention to the classic problems and methods of
quantum mechanics.  Precise thin-film deposition techniques allow of immediate
tests of the theory, as for example, the experimental observation of localized
``surface states'' in AlGaAs/GaAs superlattices with a terminating layer of
AlAs \cite{Mendez}.  Such states were predicted by Tamm in 1932 \cite{Tamm}.
These, as well as interface states (IS) in general, have since been under
permanent investigation \cite{V.-P.,KD,Tikhodeev}.  When the IS can be
associated with the periodic potential of the lattice (``Shockley states'')
\cite{buch}, they do not depend crucially on the details of the surface.
Therefore, these IS, like the emulated IS of Ref.  \cite{Mendez}, can be
studied within the envelope-function approach (EFA).

In applying the EFA to abrupt junctions, the matching conditions (MC)
connecting the wave functions across the interface have to be established.  It
is known that in a heterojunction with band inversion, IS occur as an intrinsic
property of the contact regardless of the precise type of MC \cite{V.-P.,KD}.
In the more usual case of a heterojunction of two semiconductors with the same
band symmetry, the role of the MC is more critical.  The usual continuity
conditions do not give rise to localized states, as is intuitively obvious when
applying conventional quantum mechanics to a band-edge profile as shown in the
inset in Fig.  1a.  However, it was found for a model superlattice
\cite{Tikhodeev} that IS do occur, provided the wave functions are taken to be
discontinuous at the interface.  Unfortunately, within the EFA, the commonly
used single-band description does not allow one to determine the MC
unambiguously from the kinetic energy because the latter is not unique
\cite{Sokolov,DR}.

In the present Letter, we infer unique MC for an arbitrary abrupt semiconductor
junction within the EFA, using the two-band model \cite{KEL}.  A novel type of
{\it shallow} IS is shown to occur in a contact of two semiconductors with
bandgaps and effective masses such that their bulk dispersion curves intersect.
This means that the in\-ter\-band velocity matrix element must vary across the
junction so as to have a different bulk value on either side of the interface.
It is found that interface states of this type are also present in graded
heterojunctions.

In the two-band model, the four envelope functions combined in
the spinor $\Psi$ satisfy an effective Dirac equation of the form ($\hbar=1$)
\begin{eqnarray}
\left (  v \gamma^0 \, \mbox{\boldmath $\gamma p$}
+\gamma^0  \Delta + \varphi \right )
\Psi \, =\epsilon
\,\Psi \enspace .
\end{eqnarray}
Here, $\gamma^i\equiv (\gamma^0$, {\boldmath $\gamma$}$)$ are the Dirac
matrices, $\Delta$ is half the energy gap, $\varphi$ is the work function, and
$v$ is the interband velocity matrix element.  With constant $\Delta$,
$\varphi$, and $v$, one obtains the dispersion of a bulk semiconductor,
$\epsilon^{(\pm)}({\bf k}) = \varphi \pm (\Delta^2+v^2\,|{\bf k}|^2)^{1/2}$,
for the energy of an electron, $\epsilon^{(+)}$, and a hole, $\epsilon^{(-)}$,
respectively.  For $kv \ll \Delta$, electrons and holes have one and the same
effective mass $m^*=\Delta/v^2$, given in terms of the parameters $\Delta$ and
$v$ of the two-band model.

We now consider the case where $\Delta$, $\varphi$, and $v$ depend only on the
$z$-coordinate in the direction of growth.  Here, the momentum of free motion
in the $(x,y)$-plane, ${\bf k}_\bot$, is conserved.  The ``pseudoparity''
operator $ P= i \gamma^0 \, \gamma^3 \, (${\boldmath $\gamma$}$_\bot \, {\bf
k}_\bot)/k_\bot$, which is the analogue of the helicity operator in Dirac
theory, commutes with the Hamiltonian in (1) and has eigenvalues $\lambda = \pm
1$ \cite{I-U}.  Following Ref.  \cite{WE1}, we write (1) for each $\lambda$ in
the form of an eigenvalue problem for two scalar wave functions
$\psi_{_\lambda}$ and $\chi_{_\lambda}$ of the two hybridized bands,
\begin{eqnarray}\label{4} \left ( \begin{array}{cc}
\Delta + \varphi - \epsilon_{_{\lambda}} & v \lambda k_\bot  - v
\partial_z \\ v \lambda k_\bot + v \partial_z & -\Delta + \varphi -
\epsilon_{_{\lambda}} \end{array} \right )
\left (
\begin{array}{c} \psi_{_\lambda}\\ \chi_{_\lambda}
\end{array} \right ) = 0\enspace .
\end{eqnarray}
For a graded system with a nonconstant $v$, the kinetic term $v\,\partial_z$ in
(2) must be made Hermitian by replacing it with its symmetrized form
$v(z)\,\partial_z+v^\prime(z)/2$.

In the limit of an abrupt heterojunction with discontinuous $v(z)$, the term
$v^\prime(z)/2$ gives a $\delta$-function contribution in (1) and (2).  For a
consistent description, we introduce functions $f_{_{\lambda}}$ and
$g_{_{\lambda}}$ by writing $\psi_{_{\lambda}} = f_{_{\lambda}}/\sqrt{v}$ and
$\chi_{_{\lambda}} = g_{_{\lambda}}/\sqrt{v}$, so that $\left\lbrack
v(z)\partial_z+ v^\prime(z)/2 \right\rbrack
f_{_{\lambda}}/\sqrt{v}=\sqrt{v}f_{_{\lambda}}^\prime$, and similarly for
$g_{_{\lambda}}$.  It emerges that the two-band equation for $f_{_{\lambda}}$,
$g_{_{\lambda}}$ with $z$-dependent $v$ has again the form (2), but the new
functions $f_{_{\lambda}}$, $g_{_{\lambda}}$ are continuous at the interface
$z=0$.  The MC for the physical envelope function, satisfying the Hermitian
version of (2), thus become discontinuous,
\begin{eqnarray}
\psi_{_{\lambda}}\sqrt{v}|_{-0}=\psi_{_{\lambda}}\sqrt{v}|_{+0}
\enspace , \ \ \ \ \chi_{_{\lambda}}\sqrt{v}|_{-0}=\chi_{_{\lambda}}
\sqrt{v}|_{+0} \enspace .
\end{eqnarray}
This type of MC departs from the widely used MC of a continuous envelope
function across the interface \cite{Bastard,bur88}, established for
heterojunctions with identical Bloch functions on either side, for which $v$
is constant. The assumption of constant $v^2=\Delta/m^*$ is rather well
justified for most narrow-band heterojunctions of III-V semiconductors.
However, for most IV-VI and some III-V heterojunctions, $v$ does change
appreciably across the interface \cite{Man1}, and the MC (3) should be used.
A similar situation is discussed in Ref. [5] for a one-dimensional ($k_\bot=0$)
superlattice. In contrast to that work we consider the full three-dimensional
case and derive the in-plane dispersion curves for the IS.

We study an abrupt contact of two homogeneous semiconductors with the
parameters $\Delta_i$, $v_i$, and $\varphi_i$ ($i=1$, $2$ for $z<0$, $z>0$,
respectively), where $\varphi_1 = 0$, $\varphi_2 = -V$ (cf. the inset in Fig.
1a). The IS wave function has the form $\exp(+\kappa_1z)$, $\exp(-\kappa_2z)$
for $z<0$, $z>0$, respectively, where $\kappa_i=\kappa_i(\epsilon, k_\bot)=
+[k^2_\bot+(\Delta^2_i-(\epsilon-\varphi_i)^2)/v^2_i]^{1/2}$ ($i=1$, $2$).
The in-plane ($k_z=0$) bulk dispersions $\epsilon^{(\pm)}_i(k_\bot)$
are found by setting  $\kappa_i=0$.

The bound-state eigenvalues are obtained from (2) with the aid of the MC (3),
which leads to the eigenvalue equation
\begin{eqnarray}
v_1(\lambda k_\bot+\kappa_1)/(\epsilon+\Delta_1)=
v_2(\lambda k_\bot-\kappa_2)/(\epsilon+V+\Delta_2) \enspace .
\end{eqnarray}
Equation (4) can also be written in the form $\kappa_1 \kappa_2=
\epsilon_{12}(\epsilon, k_\bot)$, where we have defined
$\epsilon_{12}(\epsilon, k_\bot)=(\epsilon^2 +\epsilon V -\Delta_1
\Delta_2)/(v_1 v_2) - k^2_\bot$. This requires $\epsilon_{12}(\epsilon,
k_\bot) > 0$, which implies $|\epsilon| > |\epsilon_0(k_\bot)|$, where
$\epsilon_0(k_\bot)$ solves $\epsilon_{12}(\epsilon_0, k_\bot) = 0$. From (4)
one obtains the IS energy $\epsilon_\lambda$ as a function of the momentum
$k_\bot$,
\begin{eqnarray} \epsilon_\lambda(k_\bot)=[-V(k^2_\bot v_1
v_-+\Delta_1 \Delta_-) - \lambda k_\bot D  (k^2_\bot v^2_-+\Delta^2_-
-V^2)^{1/2}]/(k^2_\bot  v^2_- +\Delta^2_-) \enspace ,
\end{eqnarray}
where $v_-=v_1-v_2$,  $\Delta_-=\Delta_1-\Delta_2$, and
$D=\Delta_2 v_1 - \Delta_1 v_2$. The localized states exist in a region of
momenta  ${\rm min} \{k^{(\lambda)}_{\bot i}\}<k_\bot<{\rm max}
\{k^{(\lambda)}_{\bot i}\}$, where the end points $k^{(\lambda)}_{\bot i}$
are determined by the intersection of the curves $\epsilon_0(k_\bot)$ with
the bulk dispersion curves $\epsilon_{1, 2}(k_\bot)$, respectively,
\begin{eqnarray} k^{(\lambda) 2}_{\bot i}=[V^2v_i-2v_-\Delta_i\Delta_-
+{\rm sgn} (\lambda k_\bot) V(4\Delta_iv_-D+V^2v^2_i)^{1/2}]/(2v_iv^2_-)
\enspace , \ \enspace i=1, 2 \enspace .
\end{eqnarray}

Equation (5) has been obtained without imposing any restrictive conditions
on the parameters $v_i$, $\Delta_i$, and $V$. If $v_-=0$, one recovers from (5)
the results of Refs. \cite{V.-P.,KD}, viz. that IS with linear
dispersion exist in the band-inverted case ($\Delta_1\Delta_2<0$).
Moreover, relation (6) for the end points reduces to $k^{(\lambda) 2}_{\bot
i}=(\Delta^2_--V^2)\Delta^2_i/V^2v^2$, showing that the IS exist in a finite
range of $k_\bot$ for nonzero work function offset $V\neq 0$ (cf. \cite{KD}).

On the other hand, if $v_-\neq0$, Eq. (5) describes a novel type of IS which
has a nonlinear dispersion and which, in particular, occurs in junctions {\it
without} band inversion. For this, it is necessary that a crossing of the bulk
dispersion curves occurs. This is seen immediately for an unshifted
heterojunction with $V=0$ where (6) has a real solution $k^{(\lambda)}_{\bot
i}$ only if  $\Delta_-v_-<0$ ($\Delta_{1}<\Delta_{2}$, $v_1>v_2$ or
$\Delta_{1}>\Delta_{2}$,  $v_1<v_2$), i.e. if the bulk dispersion curves
$\epsilon^{(\pm)}_i(k_\bot)$  intersect.

The behavior of the IS as a function of the offset $V$ is illustrated in
Fig. 1  for the following parameters: $\Delta_1=240$ meV, $\Delta_2=300$ meV,
$v_1=2 \times 10^8$ cm s$^{-1}$, $v_2=10^8$ cm s$^{-1}$. The IS lie in the
vicinity of the crossing  point of the bulk dispersions. It is there where
their wave functions are well localized ($\kappa_i$ ``large''). Moving away
from this point the energy of the IS approaches the bulk dispersion curves
so that the wave functions become less localized. We see that the IS are
{\it shallow} : for small values of $V$ the IS energies
$\epsilon_{-1}(k_\bot)$ and $\epsilon_{+1}(k_\bot)$ lie near the bulk
dispersions. With increasing $V$ the region $k^{(-)}_{\bot 1}<
k_\bot<k^{(-)}_{\bot 2}$
widens  whereas  the region $k^{(+)}_{\bot 1}<k_\bot<k^{(+)}_{\bot 2}$
narrows. For $V<|\Delta_-|$, the heterojunction is of type I, with two IS
having opposite $\lambda$ for given $k_\bot$, which becomes type II for
$V>|\Delta_-|$, with only the IS near the valence band remaining.

The analytic results for an abrupt junction remain virtually the same
for graded junctions. Using for $v(z)$ and $\Delta(z)$ a simple representation
involving a smoothness parameter $d$, we find by numerical calculation that
the energy of the IS changes by no more than $\pm 0.5$ meV as a function of
$d$ up to values as large as $d \approx 3 \times 10^{-6}$ cm ($\kappa_i \, d
\approx 1$, $\kappa_i \approx k_\bot$), and thus essentially coincides with
the energy obtained for the abrupt contact.

Our case is a  typical example of  weak localization (characterized by
the weak dependence of the IS on  $d$ for $\kappa^{-1}_i \gg d$), similar to
what is found in both the one-band  and two-band approximations \cite{V.-P.}.
This is seen from an analysis of the graded junction, which also throws some
light on the physical meaning of the MC in the abrupt case. Eliminating $g$
from (2) (written for the functions $f$, $g$ with $\lambda=+1$), one arrives at
a Klein-Gordon-type equation  for $f$. Using for $f$ the ansatz $f=\exp[\int
\gamma (z) {\rm d} z]$, one obtains
\begin{eqnarray} \gamma^2+(\epsilon^2-
\Delta^2)/v^2-k^2_\bot + [Q(k_\bot + \gamma)]^\prime/Q =0  \enspace ,
\end{eqnarray}
with $Q=v/(\Delta+\epsilon)$.  Away from the interface ($|z| \gtrsim d$), the
last term on the LHS of (7) vanishes, so that there $\gamma \approx \kappa_1$,
$-\kappa_2$.  In the interface region ($|z| \lesssim d$), $Q^\prime$ and
$\gamma^\prime$ ($\approx \kappa/d$) predominate over the other terms ($\approx
\kappa^2$), so that $Q(k_\bot+\gamma)|_{z_1}=Q(k_\bot+ \gamma)|_{z_2}$, where
$z_1<0$, $z_2>0$ ($|z_i| \lesssim d$).  Replacing $Q$ and $\gamma$ with their
bulk values, $Q(z_i) \rightarrow v_i/(\epsilon+\Delta_i)$, $\gamma(z_1)
\rightarrow \kappa_1$, $\gamma(z_2) \rightarrow -\kappa_2$, we recover Eq.  (4)
(with $V=0$).  The last term on the LHS of (7) acts like a potential which
turns around the sign of the wave vector $\gamma(z)$ as $z$ passes through the
interface.  In the abrupt case its role is taken over by the MC.

From our analysis, the following physical picture emerges. The free motion in
the interface plane with wave number $k_\bot \neq 0$ gives rise to binding in
the $z$-direction. The energies of the electron-like IS lie above the bulk
conduction band edges $\Delta_1$ and $\Delta_2-V$ (and analogously for the
hole-like IS). Thus, the IS are embedded in the continuum of the carriers;
nevertheless, they lie in the gap if one compares their energies with those
of the two bulk dispersions for the same value of $k_\bot$. The IS originate
from the intersecting bulk states, they are ``pushed'' down into the bandgap
owing to the negative energy contribution of the (quantized) motion in the
$z$-direction.

The principal feature of the eigenvalue equation (4) is that it contains
$k_\bot$ explicitly, so that in a range of values of $k_\bot$, the two
exponentially decaying left and right solutions can be matched at $z \approx 0$
when both $\kappa_i$ are small.  This implies a change of sign of the
logarithmic derivative.  Formulations which do not take this into account
cannot describe our IS \cite{Bastard}.  The region $k^{(\lambda)}_{\bot
1}<k_\bot<k^{(\lambda)}_{\bot 2}$ in which the IS exist could be quite broad
for IV-VI semiconductors as $v$ varies here over a wide range.  The III-V
semiconductors are generally well described within the approach of
\cite{Bastard}.  However, in extreme cases the values of $v$ differ by as much
as $0.2 \times 10^8$ cm s$^{-1}$ for materials of this group (cf., e.g., Ref.
[14], ch.  II, Table III), allowing IS in a small range of $k_\bot$.

The essential condition for the existence of the IS in
heterojunctions without band inversion, viz. the intersection of the bulk
dispersion curves, $\Delta_-v_-<0$, holds for most narrow-band semiconductors
\cite{Man1}. This is easily understood: the larger the bandgap, the weaker the
interband  interaction, i.e. materials with large $\Delta$ have small $v$.

In summary, we have used the two-band model to derive unique matching
conditions for an abrupt heterojunction.  If the interband velocity matrix
element $v$ is different in the two semiconductors (as in most
heterojunctions), localized interface states occur, in particular in systems
without band inversion, if the bulk curves of the semiconductors intersect.
This is expected to be the normal case.  The effect of the MC is rather weak,
leading to IS which lie near the bulk dispersions.  However, in spite of this,
it should be possible to identify these states experimentally, since owing to
their localized nature, their contribution to transport and optical properties
is different from that of the bulk states.

The present model, while being not unrealistic for IV-VI semiconductors, is
simple and transparent, and should provide an ideal guideline for more
detailed theoretical treatments which are expected to lead to a more
quantitative exploration of the predicted type of state.  Moreover, beyond the
present context of the envelope function description, our results represent an
important and novel contribution to quantum mechanics in general, as they
demonstrate the existence of localized interface states in a situation where,
without the manifest occurrence of a binding potential, localization is
seemingly impossible.
\begin{center}
***
\end{center}
This material is based upon work supported by the U.S. Civilian Research
and Development Foundation under Award No.  RP1-273. The present work was
supported in part by the Russian Foundation for the Fundamental Research
(Projects No. 96-02-16701, 97-02-16346, and  97-02-16042).

\bigskip
\newpage
\noindent
{\Large \bf Figure captions}

\smallskip \noindent Figure 1. Dispersion $\epsilon_\lambda(k_\bot)$ of an abrupt
heterojunction for several values of $V$ (heavy curves; for parameters, see text).  Dotted
curves:  bulk dispersions $\epsilon^{(\pm)}_{1, 2}(k_\bot)$.  The regions (with end points
$k^{(\lambda)}_{\bot i}$) where localized IS occur are defined by the condition
$|\epsilon_\lambda(k_\bot)|>|\epsilon_0(k_\bot)|$, where $\epsilon_0(k_\bot)$ is represented
by the thin curves.  It is understood that the figures are continued symmetrically across
$k_\bot=0$ with a reversal of sign of $\lambda$. Inset in a):  band-edge profile of the
heterojunction; the energy gap is shaded. \newpage \begin{figure}[ht] \begin{center}
  \unitlength1cm
  \begin{minipage}[t]{12cm} \hspace*{-2cm}
  \begin{picture}(6,6)
     \put(14.5,9.)
    {\includegraphics{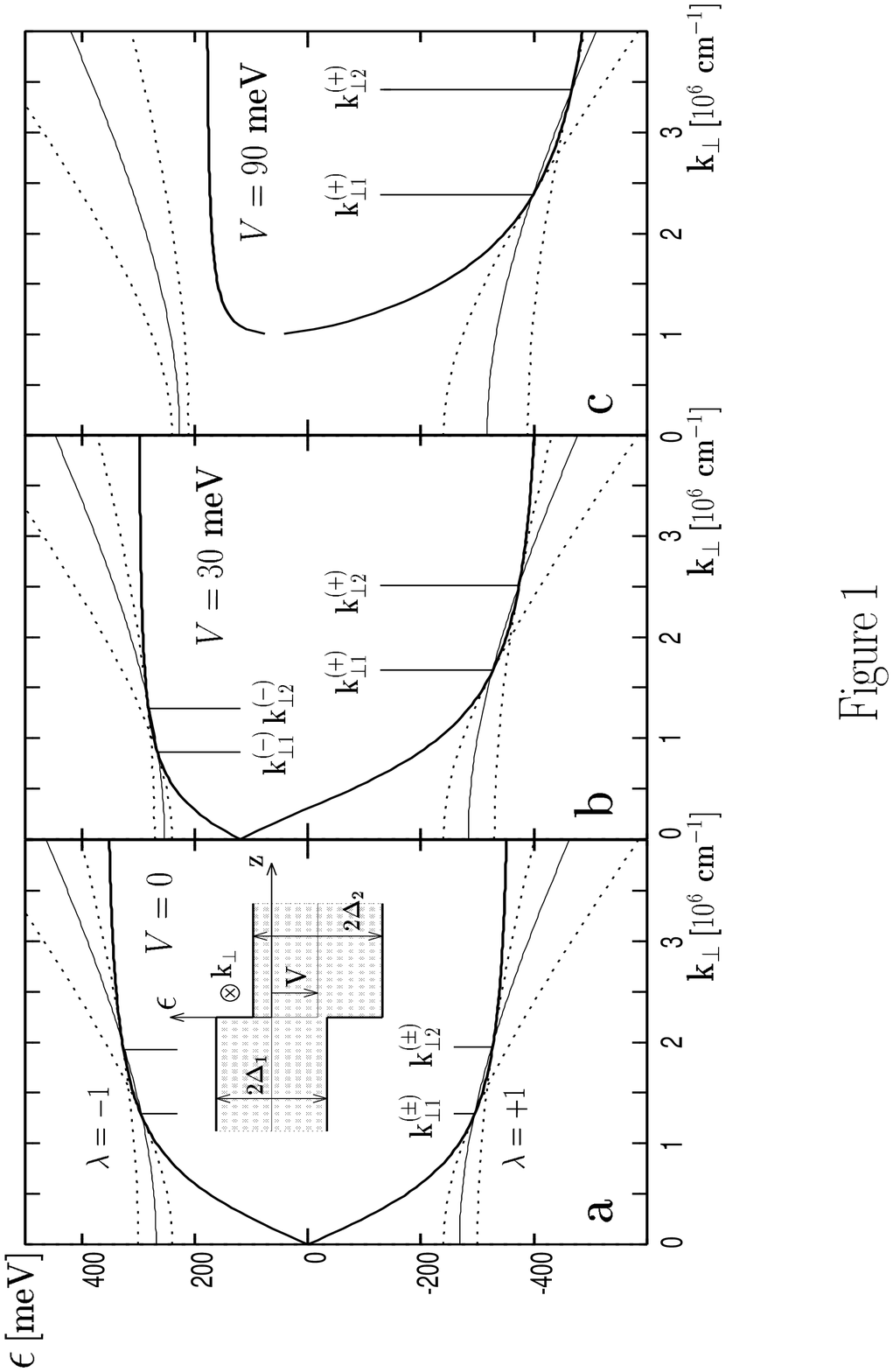}}
 \end{picture}
  \end{minipage} \end{center}
\end{figure}                                                                   
\end{document}